\documentclass[12pt,english,preprint,superscriptaddress,showpacs,nofootinbib,
nobibnotes]{iopart}

\usepackage[T1]{fontenc}
\usepackage[utf8x]{inputenc}
\usepackage{cite}
\usepackage{graphicx}
\usepackage{amssymb}

 \usepackage{amsfonts}
 \usepackage{dsfont}
 \usepackage{epsfig}
 \usepackage{subfigure}
 \usepackage{graphicx}
 \usepackage{caption}
 \usepackage{float}

\makeatletter
\@ifundefined{textcolor}{}
{%
 \definecolor{BLACK}{gray}{0}
 \definecolor{WHITE}{gray}{1}
 \definecolor{RED}{rgb}{1,0,0}
 \definecolor{GREEN}{rgb}{0,1,0}
 \definecolor{BLUE}{rgb}{0,0,1}
 \definecolor{CYAN}{cmyk}{1,0,0,0}
 \definecolor{MAGENTA}{cmyk}{0,1,0,0}
 \definecolor{YELLOW}{cmyk}{0,0,1,0}
 }

\makeatother

\usepackage{babel}

\begin{document}

\title[Non-Equilibrium Quantum Dynamics of Ultra-Cold Atomic
Mixtures]{Non-Equilibrium Quantum
Dynamics of Ultra-Cold Atomic Mixtures:
the Multi-Layer Multi-Configuration Time-Dependent Hartree Method for Bosons}

\author{Sven Kr\"onke$^{1,3}$, Lushuai Cao$^{1,3}$, Oriol Vendrell$^{2,3}$,
Peter Schmelcher$^{1,3}$}
\address{$^1$ Zentrum f\"{u}r Optische Quantentechnologien, Universit\"{a}t
Hamburg, Luruper Chaussee 149, D-22761 Hamburg, Germany}
\address{$^2$ Center for Free-Electron Laser Science, DESY, Notkestrasse 85, 
D-22607 Hamburg, Germany}
\address{$^3$ The Hamburg Centre for Ultrafast Imaging,
Luruper Chaussee 149,
D-22761 Hamburg, Germany}
\ead{\mailto{sven.kroenke@physnet.uni-hamburg.de}, 
\mailto{lcao@physnet.uni-hamburg.de},
\mailto{oriol.vendrell@cfel.de}, 
\mailto{pschmelc@physnet.uni-hamburg.de}}

\begin{abstract}
We develop and apply the multi-layer multi-configuration
time-dependent Hartree method for bosons, which represents an
{\it ab initio} method for investigating the non-equilibrium quantum dynamics of
multi-species bosonic
systems. Its multi-layer feature allows for tailoring the
wave
function ansatz in order to describe intra- and inter-species correlations
accurately and efficiently. To demonstrate the beneficial scaling and the
efficiency of the method, we explore the correlated tunneling dynamics of two
species with repulsive intra- and inter-species interactions, to which a third
species with vanishing intra-species interaction is weakly coupled. The
population imbalances of the first two species can feature a temporal
equilibration and their time-evolution significantly depends on the
coupling to the third species. Bosons of the first and of the second
species exhibit a bunching tendency, whose strength can be influenced by
their coupling to the third species.
\end{abstract}


\maketitle

%%%%%%%%%%%%%%%%%%%%%%%%%%%%%%%%%%%%%%%%%%%%%%%%%%%%%%%%%%%%%%%%%%%%%%%%%%%%%%%%
%%%%%%%%%%%%%%%%%%%%%%%%%%%%%%%%%%%%%%%%%%%%%%%%%%%%%%%%%%%%%%%%%%%%%%%%%%%%%%%%
\section{Introduction}\label{sec_intro}
Due to the high degree of controllability and isolatedness, trapped
ultra-cold
atoms serve as an ideal system for observing many-body quantum phenomena
\cite{many_body_physics_ultrac_atoms_zwerger_dalibard_bloch} and can even
be employed to simulate quantum systems of quite a broad physical context
\cite{quantum_simulators_with_ultracold_quantum_gases_Bloch2012}.
In particular, there is a growing
interest  in the meanwhile accessible regime where a mean-field description
\cite{Bose-Einstein_Condensation_Dilute_Gases_Pethick_Smith_2008,
Pitaevskii_Stringari_Bose-Einstein_Condensation2003} as
given by the Gross-Pitaevskii equation fails. Such states
can be realized in
e.g. optical lattices
\cite{quantum_phase_transition_SF_to_MI_Greiner_Nature2002}. Feshbach
\cite{Feshbach_resonances_ultracold_gase_Grimm_2010} or
confinement induced resonances
\cite{Olshanii_PRL98_quasi_1d_scattering,
observation_of_1d_TG_gas_Kinoshita_Science2004,
confinement_induced_molecules_in_1d_fermigas_Esslinger_PRL2005} can be employed
to tune
the inter-atomic interaction strength. In particular, quasi-one-dimensional
trapping geometries can enhance correlation effects in the strong interaction
regime, leading to fascinating novel phases
\cite{girardeau_bose_fermi_map_1960,
observation_of_1d_TG_gas_Kinoshita_Science2004,
TG-gs_of_atoms_in_opt_lattice_Bloch_Nature2004,
realization_of_sTG_gas_Haller_Science2009} and quantum phase transitions
\cite{pinning_quantum_phase_transisition_Haller2010}. As the mean-field theory
becomes exact for weak interactions and large particle numbers
\cite{proof_BEC_for_dilute_trapped_gases_Lieb_Seiringer,
Bose-Einstein_Condensation_Dilute_Gases_Pethick_Smith_2008,
Pitaevskii_Stringari_Bose-Einstein_Condensation2003}, beyond
mean-field physics can also be expected for small ensembles and finite
interaction strengths. The latter regime is experimentally explored in
e.g. arrays of decoupled one-dimensional
tubes typically containing two to 60 atoms
\cite{pinning_quantum_phase_transisition_Haller2010}. Therefore, the
transition from the few- to many-body behaviour is, in particular for the
strongly correlated quantum dynamics, a subject of immediate interest.

Moreover it is meanwhile also experimentally routinely achievable to trap
and manipulate different
components or
species\footnote{In the following, we will use the term {\it species}
irrespectively of whether it refers to different elements, isotopes or
internal states of an isotope.}, which allows for studying distinguishable
subsystems with indistinguishable constituents. Such mixtures can be
realized e.g. by preparing alkali atoms in different hyperfine states
\cite{production_of_2_overlapping_BECs_symopathic_cooling_WiemanPRL1997} or by
trapping different elements
\cite{bec_of_Potassium_by_sympathic_cooling_Inguscio_Science2001}. Due to
the interplay between the intra- and inter-species interaction strengths, these
systems show a number of intriguing features such as phase separation
\cite{dynamics_of_comp_separation_in_binary_BEC_Cornell_PRL1998} including
symbiotic excitations like interlacing vortex lattices with mutually filled
cores \cite{vortex_lattice_dynamics_in_rotating_spinor_BEC_Cornell_PRL2004} and
dark-bright solitons
\cite{oscillations_and_int_of_dark_and_darkbright_solitons_Becker_Nature2008},
spin-charge separation
\cite{spin-charge_sep_in_2comp_Bosegase_Schollwoeck_PRA2008}, various tunneling
effects
\cite{dyn_of_2_spec_BEC_in_double_well_Pindzola_PRA2009,
josephson_osc_in_spinor_BEC_Sanpera_PRA2009,
symmetry_breaking_and_restoring_in_mixture_BEC_double_well_Clark_PRA2009,
quantum_bose_josephson_junct_with_binary_BEC_Citro_JPB2010,
material_barrier_tunneling_1d_fewboson_mixtures_Pflanzer_Zoellner,
interspecies_tunneling_1d_bose_mixtures_Pflanzer_Zoellner,
correl_mixture_ tunneling_double_well_Chatterjee_PRA2012,
impact_of_spatial_correl_on_tunneling_mixture_Capo2012}, collective
excitations
\cite{collect_osc_of_2_colliding_BECs_Inguscio_PRL2000,
dyn_of_2_colliding_BECs_Minardi_PRA2000} and counterflow and paired
superfluidity
\cite{counterflow_SF_of_2spec_commens_lattice_Svistunov_PRL2003,
counterflow_and_paird_S F_in_1d_Bose_mixtures_Clark_PRA2009,
detecting_paired_counterflow_SF_via_dipole_osc_Clark_PRA2011}. The purpose of
the present work is to develop a broadly applicable and efficient {\it ab
initio} method for the quantum dynamics of such mixtures in order to explore
the fundamental dynamical processes in trapped ultra-cold multi-species setups
and to study the few- to many-body transition.

Simulating the quantum dynamics of an interacting many-body system, however,
is a tough task in general due to the exponential\footnote{For distinguishable
particles. Indistinguishable particles lead to a binomial scaling.} scaling of
the state space with the number of particles. 
Besides e.g. the time-dependent density matrix renormalization group
approach \cite{real_time_evo_using_DMRG_White_Feiguin_PRL2004,
td_DMRG_using_adaptive_eff_Hilbert_spaces_Vidal2004,
td-DMRG_methods_Schollwoeck2004, DMRG_at_age_of_MPS_Schollwoeck2010}, a
promising concept to soften
this scaling is based on a many-body wave function expansion with respect to a
time-dependent, with the system comoving basis. This idea has been
incorporated in the multi-configuration time-dependent Hartree method (MCTDH)
\cite{mctdh_approach_Meyer_Manthe_Cederbaum,MCTDH_BJMW2000}. Being based
on time-dependent Hartree products as the many-body basis, MCTDH is designed
for distinguishable particles, but has also been applied to bosonic few-body
systems (e.g.
\cite{few_boson_dyn_doubl_wells_Zoellner,
impact_of_spatial_correl_on_tunneling_mixture_Capo2012}). Later the MCTDH
theory has been generalized and extended in several ways: There is the
multi-layer MCTDH (ML-MCTDH) method
\cite{Multilayer_formulation_Wang_Thoss_2003,
Multilayer_multiconfig_Manthe_2008,
 ML_implementation_Vendrell_Meyer_JChemPhys_2011}, which takes correlations
between various subsystems into account and is thus
particularly suitable for system-bath problems with distinguishable degrees
of freedom (e.g. \cite{wang_shao12}). Taking
the fermionic or bosonic
particle exchange \mbox{(anti-)} symmetry in the time-dependent many-body basis
into account, MCTDH has been specialized to
treat larger fermionic (MCTDHF)
\cite{MCTDHF_approach_to_multi_electr_dynamics_in_laser_fields_Scrinzi2003} or
bosonic systems (MCTDHB)
\cite{
role_of_excited_states_in_splitting_of_BEC_by_time_dep_barrier_Steltsov_PRL2007,
MCTDHB_many_body_dynacmics_of_bosonic_systems_phys_rev_a_Alon_Streltsov_Cederbau
m}. Furthermore, a direct
extension of MCTDHB and MCTDHF to treat bose-bose, bose-fermi and fermi-fermi
mixtures has been
developed \cite{MCTDHX_recursively_Cederbaum2012}, including the possibility
of particle-conversions
\cite{MCTDHB_with_particle_conversion_Alon_Streltsov_Cederbaum_PRA2009}. An
alternative approach to systems of indistinguishable particles is the so-called
ML-MCTDH method in second-quantization representation, which employs
the factorization of the many-body Hilbert space into a direct product of
Fock spaces
\cite{MLMCTDH_in_2nd_quantization_representation_Wang_Thoss_Chem_Phys_2009}.

In this work, we derive and apply a novel {\it ab initio} approach to the
non-equilibrium dynamics of ultra-cold correlated bosonic mixtures, which takes
all correlations of the
many-body system into account. We call this method the {\it multi-layer
multi-configuration time-dependent
Hartree method for bosons} (\mbox{ML-MCTDHB}). The
multi-layer
structure of our many-body wave function ansatz allows us to adapt our many-body
basis to system specific inter- and intra-species correlations, which leads to a
beneficial scaling. Moreover, the bosonic exchange symmetry is directly employed
for an efficient treatment of the indistinguishable bosonic subsystems. We apply
ML-MCTDHB to simulate the correlated tunneling dynamics of a
mixture of three bosonic species in a double well trap. It is shown that the
dynamics of the population imbalances of the species significantly differ
for ultra-weak and vanishing inter-species interaction strengths. In
particular, bosons of different kind show a bunching tendency
and the inter-species interaction strengths allow for tuning these correlations
up to a certain degree.  

This paper is organized as follows: In section \ref{sec_method}, the
derivation and properties of the \mbox{ML-MCTDHB} method for bosonic
mixtures are presented. \mbox{ML-MCTDHB} is then applied to simulate the
complex tunneling behaviour of a mixture of three bosonic species in
section \ref{sec_appl}. Finally, we summarize our results and embed the
presented \mbox{ML-MCTDHB} theory for mixtures into a more general framework in
section
\ref{sec_concl}. 
%%%%%%%%%%%%%%%%%%%%%%%%%%%%%%%%%%%%%%%%%%%%%%%%%%%%%%%%%%%%%%%%%%%%%%%%%%%%%%%%
%%%%%%%%%%%%%%%%%%%%%%%%%%%%%%%%%%%%%%%%%%%%%%%%%%%%%%%%%%%%%%%%%%%%%%%%%%%%%%%%
\section{The \mbox{ML-MCTDHB} method}\label{sec_method}
Let us consider an ensemble of $S$ bosonic species. In the
ultra-cold
regime, the interaction between neutral atoms can be modelled
by a contact interaction \cite{Huang_Yang_paper_PhysRev1957,
Bose-Einstein_Condensation_Dilute_Gases_Pethick_Smith_2008,
Pitaevskii_Stringari_Bose-Einstein_Condensation2003}. For simplicity, we
restrict ourselves to
one-dimensional settings, which can be prepared by energetically freezing out
the transversal degrees of freedom
\cite{Bose-Einstein_Condensation_Dilute_Gases_Pethick_Smith_2008}. The
Hamiltonian of such a mixture
with $N_\sigma$ bosons of species $\sigma=1,...,S$
reads:
\begin{equation}\label{hamil_tot}
 \hat H =\sum_{\sigma=1}^S \left(\hat \mathcal{H}_\sigma+\hat V_\sigma\right)
	 +\sum_{1\leq \sigma<\sigma'\leq S} \hat W_{\sigma\sigma'}.
\end{equation}
Here, $\hat \mathcal{H}_\sigma$ denotes the one-body Hamiltonian of the species
$\sigma$
containing a in general species dependent trapping potential $U_\sigma$:
\begin{equation}\label{hamil_1b}
 \hat \mathcal{H}_\sigma =%\sum_{i=1}^{N_\sigma} \hat h^\sigma_i=
\sum_{i=1}^{N_\sigma}
\left(
    \frac{\hat p^\sigma_i\phantom{)\!\!}^2}{2m^\sigma}+
	  U_\sigma(\hat x_i^\sigma)\right),      
\end{equation}
and $\hat V_\sigma$, $\hat W_{\sigma\sigma'}$ refer to the intra-species
interaction of species $\sigma$ and to the inter-species interaction between
$\sigma$ and $\sigma'$ bosons, respectively:
\begin{eqnarray}\label{hamil_intra}
 \hat V_\sigma =g_\sigma\sum_{1\leq i<j \leq N_\sigma} \delta(\hat x_i^\sigma-
		\hat x_j^\sigma),\\\label{hamil_inter}
 \hat W_{\sigma\sigma'} = g_{\sigma\sigma'}
		\sum_{i=1}^{N_\sigma}\sum_{j=1}^{N_{\sigma'}}
		\delta(\hat x_i^\sigma-\hat x_j^{\sigma'}).
\end{eqnarray}
Please note that the intra- and inter-species interaction strengths $g_\sigma$,
$g_{\sigma\sigma'}$ have to be properly renormalized with respect to their 3d
values as a consequence of dimensional reduction
\cite{Olshanii_PRL98_quasi_1d_scattering}. We remark that the Hamiltonian may be
explicitly time-dependent for studying driven systems.

\subsection{Wavefunction ansatz}\label{ssec_ansatz}
The \mbox{ML-MCTDHB} method is an {\it ab initio} approach to the time-dependent
Schr\"odinger equation for systems like (\ref{hamil_tot}). To reduce the
number of basis states necessary for a fair representation of the total
wave function $|\Psi(t)\rangle$, we employ a time-dependent, with the system
comoving basis and restrict ourselves to the following class of ansatzes: For
each species $\sigma$, we take $M_\sigma$ time-dependent orthonormal species
states
$|\psi^{(\sigma)}_i(t)\rangle$ ($i=1,...,M_\sigma$), i.e. states of all the
$N_\sigma$ bosons of species $\sigma$, into account. Due to the
distinguishability of bosons of different species, the total wave function is
expanded in terms of Hartree products of these many-body states:
\begin{equation}\label{exp_top}
 |\Psi(t)\rangle = \sum_{i_1=1}^{M_1}...\sum_{i_S=1}^{M_S} A_{i_1,...,i_S}(t)\;
  |\psi^{(1)}_{i_1}(t)\rangle ... |\psi^{(S)}_{i_S}(t)\rangle.
\end{equation}
Each species state $|\psi^{(\sigma)}_i(t)\rangle$ refers to a system of
$N_\sigma$ indistinguishable bosons and should therefore be expanded in terms
of bosonic number states $|\vec n\rangle^\sigma_t$:
\begin{equation}\label{exp_spec}
  |\psi^{(\sigma)}_{i}(t)\rangle = \sum_{\vec n|N_\sigma} 
      C^\sigma_{i;\vec n}(t)\;|\vec n\rangle^\sigma_t, 
\end{equation}
where we allow each $\sigma$ boson to occupy $m_\sigma$ time-dependent single
particle
functions (SPFs) $|\phi^{(\sigma)}_j(t)\rangle$, indicated by the
time-dependence of bosonic number states $|\vec n\rangle^\sigma_t$. The integer
vector $\vec
n=(n_1,...,n_{m_\sigma})$ contains the occupation number $n_j$ of the $j$-th
SPFs such that all $n_j$'s sum up to $N_\sigma$, indicated by the symbol
``$\vec n|N_\sigma$'' in the summation.

Summarizing, our wave function ansatz consists of three layers: The expansion
coefficients $A_{i_1,...,i_S}(t)$ form the top layer. Then we have the 
$C^\sigma_{i;\vec n}(t)$'s on the species layer which allow the species states
to move with the
system and, finally on the particle layer, the SPFs
$|\phi^{(\sigma)}_j(t)\rangle$ allow for rotations of the single particle
basis. It is crucial to notice that, in contrast to the standard method for
solving the time-dependent Schr\"odinger equation by propagating expansion
coefficients while keeping the basis time-independent, ML-MCTDHB is based on an
expansion with respect to a comoving basis with a two-fold time-dependence in
terms of the species states $|\psi^{(\sigma)}_{i}(t)\rangle$ and the SPFs
$|\phi^{(\sigma)}_j(t)\rangle$. This two-fold time-dependence allows for
significantly reducing the number of basis states leading to a very efficient
algorithm. Please also note that our \mbox{ML-MCTDHB} approach to mixtures
conceptually differs from ML-MCTDH in second-quantization representation
\cite{MLMCTDH_in_2nd_quantization_representation_Wang_Thoss_Chem_Phys_2009} by
the facts that we only employ two layers, one for the whole species and one for
the single bosons, but allow for a time-dependent single
particle basis.

Having the number of grid points for representing the SPFs fixed, the
numbers of species states $M_\sigma$ and SPFs $m_\sigma$ serve
as numerical control parameters: Taking $m_\sigma$ to be equal to the number of
grid points and $M_\sigma$ equal to the number of number state configurations,
i.e. $(N_\sigma+m_\sigma-1)!/[N_\sigma!(m_\sigma - 1)!]$, the ansatz
(\ref{exp_top},\ref{exp_spec}) proves to be numerically exact. Opposite to this
full CI limit, the choice $m_\sigma=M_\sigma=1$ leads to the mean-field or
Gross-Pitaevskii approximation
\cite{Bose-Einstein_Condensation_Dilute_Gases_Pethick_Smith_2008,
Pitaevskii_Stringari_Bose-Einstein_Condensation2003}. In between
these two limiting case, any choice with $m_\sigma$ equal to or smaller than the
number of
grid points and $M_\sigma\leq(N_\sigma+m_\sigma-1)!/[N_\sigma!(m_\sigma - 1)!]$
is possible, which allows us to adapt our ansatz to system specific intra- and
inter-species correlations. If, for instance, the inter-species interactions are
relatively weak compared to the intra-species interactions, a ``species
mean-field'' ansatz with $M_\sigma=1$ but $m_\sigma>1$ might be sufficient.

\subsection{Equations of motion}\label{ssec_eqm}
Our final task is to find appropriate equations of motion for the ansatz
constituents $A_{i_1,...,i_S}(t)$, $C^\sigma_{i;\vec n}(t)$ and
$|\phi^{(\sigma)}_j(t)\rangle$, whose time dependence we will omit in the
notation from now on. In order
to find the variationally optimal wave function
$|\Psi(t)\rangle$ within  our class of ansatzes for given $M_\sigma$,
$m_\sigma$, we can employ the McLachlan variational principle, which enforces
the minimization of the error of our equations of motion with respect to the
exact Schr\"odinger equation \cite{mc_lachlan_1963}. In practice, however, it
is easier to work with the Dirac-Frenkel variational principle $\langle
\delta\Psi|(i\partial_t-\hat H)|\Psi\rangle=0$ with $|\delta\Psi\rangle$ being
a variation within our ansatz class ($\hbar\equiv1$)
\cite{dirac_variational,frenkel_variational}, which turns out to be equivalent
to McLachlan's variational principle on our
manifold of wave function ansatzes\cite{equivalence_of_td_var_princ}. 

The variation of the top layer coefficients $A_{i_1,...,i_S}$ gives us the
usual linear equation of motion known from matrix mechanics:
\begin{equation}\label{eqm_top}
 i\partial_t A_{i_1,...,i_S} = \sum_{j_1=1}^{M_1}...\sum_{j_S=1}^{M_S}
      \langle \psi^{(1)}_{i_1}...\,\psi^{(S)}_{i_S}|\hat H
      |\psi^{(1)}_{j_1}...\,\psi^{(S)}_{j_S}\rangle\;A_{j_1,...,j_S},
\end{equation}
where the Hamiltonian matrix with respect to Hartree products of species
states becomes time-dependent due to the coupling to the $C^\sigma_{i;\vec n}$
coefficients and to the SPFs. Its explicit form is given in
\cite{ml-mctdhb_unpub}. 

Varying the species state expansion coefficients $C^\sigma_{i;\vec n}$, we
obtain the following equations of motion on the species layer:
\begin{eqnarray}\label{eqm_spec}
\eqalign\nonumber
 i\partial_t C^\sigma_{i;\vec n} &=&  {^{\sigma\!}\langle}\vec
n|(\mathds{1}-\hat P^{1;\sigma})\;\sum_{\vec m|N_\sigma}
    \Big( \sum_{j,k=1}^{m_\sigma} {[h_\sigma]}_{jk}\,\hat a_{\sigma
j}^\dagger\hat a_{\sigma k} \;|\vec m\rangle^\sigma\;C^\sigma_{i;\vec m} + \\
    &+& \frac{1}{2}\sum_{j,k,q,p=1}^{m_\sigma}
{[v_{\sigma}]}_{jkqp}\,
\hat a_{\sigma j}^\dagger \hat a_{\sigma k}^\dagger \hat a_{\sigma q} \hat
a_{\sigma p} \;|\vec m\rangle^\sigma\;C^\sigma_{i;\vec m}+\\\nonumber
&+& \sum_{\sigma'\neq\sigma}\sum_{s,t=1}^{M_{\sigma}}
\sum_{u,v=1}^{M_{\sigma'}}\sum_{j,k=1}^{m_\sigma}
{[\eta_{1,\sigma}^{-1}]}_{is}
  \;{[ \eta_{ 2, \sigma\sigma'}]}_{sutv} 
\;{[w_{\sigma\sigma'}]}^{jk}_{uv}\;
  \,\hat a_{\sigma j}^\dagger\hat
a_{\sigma k} \;|\vec m\rangle^\sigma\;C^\sigma_{t;\vec
m}\Big),
\end{eqnarray}
where $\hat a^{(\dagger)}_{\sigma i}$ denotes the bosonic annihilation
(creation) operator corresponding to the SPF $|\phi^{(\sigma)}_i\rangle$,
obeying the canonical commutation relations $[\hat a_{\sigma i},\hat
a_{\sigma' j}]=0$ and $[\hat a_{\sigma i},\hat
a^\dagger_{\sigma' j}]=\delta_{\sigma\sigma'}\delta_{ij}$. ${[h_\sigma]}_{jk}$
and ${[v_{\sigma}]}_{jkqp}$ represent the matrix elements of the one-body
Hamiltonian and the intra-species interaction potential
with respect to the SPFs, respectively. The inter-species interaction leads to
the
mean-field matrix $[w_{\sigma\sigma'}]$ coupling both SPFs and species states.
The reduced density matrix of the species
$\sigma$ and the reduced density matrix of the subsystem constituted by the
species $\sigma$ and $\sigma'$ ($\sigma\neq\sigma'$) enter (\ref{eqm_spec}) as
$[\eta_{1,\sigma}]$ and
$[ \eta_{ 2, \sigma\sigma'}]$, respectively (cf. (\ref{rho_1spec}),
(\ref{rho_2spec})). The orthonormality of the species
states is ensured by the projector $\hat P^{1;\sigma}=\sum_{s=1}^{M_\sigma}
|\psi^{(\sigma)}_s\rangle\!\langle\psi^{(\sigma)}_s|$. Formulas for the above
ingredients are given in \ref{app_spec_layer} and an efficient scheme for
applying the annihilation and creation operators to the number states can be
found in \cite{ml-mctdhb_unpub} (see also
\cite{
general_mapping_for_bos_and_ferm_ops_in_fock_space_Streltsov_Alon_Cederbaum} in
this context).

Finally, the variation of the SPFs leads to the following non-linear
integro-differential equations:
\begin{eqnarray}\label{eqm_part}
\eqalign\nonumber
 i\partial_t|\phi^{(\sigma)}_i\rangle = (\mathds{1}-\hat P^{2;\sigma})\;
    &\Big(&\hat h_\sigma|\phi^{(\sigma)}_i\rangle + \\
    &+& \sum_{j,k,q,p=1}^{m_\sigma}
  {[\rho_{1,\sigma}^{-1}]}_{ij}
  \;{[ \rho_{ 2, \sigma\sigma}
]}_{jkqp}
  \;{[ \hat v_{\sigma}]}_{kq}\;
  |\phi^{(\sigma)}_p\rangle +\\\nonumber
&+& \sum_{\sigma'\neq\sigma}\sum_{j,q=1}^{m_\sigma}\sum_{k,p=1}^{m_{\sigma'}}
{[\rho_{1,\sigma}^{-1}]}_{ij}
  \;{[ \rho_{ 2, \sigma\sigma'}
]}_{jkqp}
  \;{[\hat
w_{\sigma\sigma'}]}_{kp}\;
  |\phi^{(\sigma)}_q\rangle	\Big).
\end{eqnarray}
Here, $[\rho_{1,\sigma}]$ denotes the
reduced density matrix of a $\sigma$ boson and $[ \rho_{ 2, \sigma\sigma}]$, 
$[\rho_{2, \sigma\sigma'}]$ ($\sigma\neq\sigma'$) refer to the reduced two-body
density matrix of
two $\sigma$ bosons, a $\sigma$ and a $\sigma'$ boson, respectively (cf.
(\ref{rho_1part})-(\ref{rho_1A1Bpart})). $\hat h_\sigma$
corresponds to the one-body Hamiltonian $\frac{\hat
p_\sigma^2}{2m^\sigma}+
	  U_\sigma(\hat x_\sigma)$ and the intra- and
inter-species interactions enter these equations of motion in the form of the
mean-field operator matrices $[\hat v_{\sigma}]$ and $[\hat
w_{\sigma\sigma'}]$, respectively. All these ingredients are explicated in
\ref{app_part_layer}.
The projector $\hat P^{2;\sigma}=\sum_{s=1}^{m_\sigma}
|\phi^{(\sigma)}_s\rangle\!\langle\phi^{(\sigma)}_s|$ again ensures the
orthonormality of the SPFs. So we have arrived at a set of highly coupled
evolution equations
(\ref{eqm_top}, \ref{eqm_spec},
\ref{eqm_part}), whose general properties we analyse in the following section.

\subsection{Properties of the \mbox{ML-MCTDHB} theory}\label{ssec_prop}
Derived from the Dirac-Frenkel variational principle, the \mbox{ML-MCTDHB}
evolution
equations preserve both norm and energy \cite{MCTDH_BJMW2000}. Moreover, one
can show that for a Hamiltonian with a (single particle) symmetry,
\mbox{ML-MCTDHB}
respects both the symmetry of the SPFs and the symmetry of the many-body
state, given that initially the SPFs and the many-body
state have a well-defined symmetry \cite{ml-mctdhb_unpub}.

In the full CI limit, i.e. $m_\sigma$ equal to the number of grid points and
$M_\sigma=(N_\sigma+m_\sigma-1)!/[N_\sigma!(m_\sigma - 1)!]$, the projectors in
(\ref{eqm_spec},\ref{eqm_part}) turn into unit operators such that both the
species states and the SPFs become time-independent. In this numerically
exact limit, \mbox{ML-MCTDHB} becomes equivalent to the standard method of
solving the
time-dependent Schr\"odinger equation by propagating only the $A$-coefficients. 
The full CI limit, however, is numerically only manageable for extremely
small
particle numbers, whereas ML-MCTDHB being based on a smaller but with the
system optimally comoving basis can treat much larger ensembles.
In the
opposite mean-field limit $m_\sigma=M_\sigma=1$, the time-dependence of the $A$-
and the $C$-coefficients is given by trivial phase factors. With all the various
reduced
density matrices being equal to the c-number one, the equations
(\ref{eqm_part}) just differ from the coupled Gross-Pitaevskii equations of the
mean-field theory for mixtures
\cite{Bose-Einstein_Condensation_Dilute_Gases_Pethick_Smith_2008,
emerg_phenom_BEC_08} by a physically irrelevant phase factor as a consequence of
the projector $\hat P^{2;\sigma}$.

A converged \mbox{ML-MCTDHB} calculation takes all correlations into account.
These
correlations can be studied by means of reduced density matrices of various
subsystems, which the \mbox{ML-MCTDHB} method provides for free.
Single-particle coherence as well as correlations between two bosons of the same
or of different species can be unravelled with the help of
$[\rho_{1,\sigma}]$, $[ \rho_{ 2, \sigma\sigma}]$ and $[\rho_{2,
\sigma\sigma'}]$.
The entropy of a species as well as correlations between two species can be
deduced from $[\eta_{1,\sigma}]$ and $[\eta_{ 2, \sigma\sigma'}]$, for example.
Moreover, an analysis of the natural populations and orbitals of various
subsystems both serves as an internal convergence check (see
below)\cite{MCTDH_BJMW2000} and can give physical insights
\cite{red_density_matrices_and_coherence_of_trapped_bosons_Sakmann_PRA2008}. 

In the case of just one species and in the full CI limit on the species
level $M_\sigma=(N_\sigma+m_\sigma-1)!/[N_\sigma!(m_\sigma - 1)!]$, the
\mbox{ML-MCTDHB} theory becomes equivalent to MCTDHB
\cite{
role_of_excited_states_in_splitting_of_BEC_by_time_dep_barrier_Steltsov_PRL2007,
MCTDHB_many_body_dynacmics_of_bosonic_systems_phys_rev_a_Alon_Streltsov_Cederbau
m} and its generalization to mixtures
\cite{MCTDHX_recursively_Cederbaum2012}, respectively. If, however, less
species states are sufficient for a converged
simulation, \mbox{ML-MCTDHB} proves to have a better scaling. With $n$ being the
number of grid points, one has to pay:
\begin{equation}
 \prod_{\sigma=1}^S M_\sigma + \sum_{\sigma=1}^S  \left ( M_\sigma\,
  {N_\sigma+m_\sigma-1 \choose m_\sigma-1} + m_\sigma\,n
\right)
\end{equation}
complex coefficients for storing a \mbox{ML-MCTDHB} wave function, which
should be compared with the costs for a corresponding MCTDHB expansion:
\begin{equation}
 \prod_{\sigma=1}^S {N_\sigma+m_\sigma-1 \choose m_\sigma-1} +
 \sum_{\sigma=1}^S m_\sigma\,n.
\end{equation}
For a detailed scaling comparison of the MCTDH type methods, we refer to
\cite{ml-mctdhb_unpub}.
%%%%%%%%%%%%%%%%%%%%%%%%%%%%%%%%%%%%%%%%%%%%%%%%%%%%%%%%%%%%%%%%%%%%%%%%%%%%%%%%
%%%%%%%%%%%%%%%%%%%%%%%%%%%%%%%%%%%%%%%%%%%%%%%%%%%%%%%%%%%%%%%%%%%%%%%%%%%%%%%%
\section{Application to correlated tunneling dynamics}\label{sec_appl}

Let us now explore the tunneling dynamics of three bosonic
species, refered to as the A, B and C species in the following, in a double
well trap. This setup both unravels interesting correlation
effects and illustrates the beneficial scaling of \mbox{ML-MCTDHB} by
introducing the
extra species layer. 

In the following, we assume that the three species are realized as different
hyperfine states of an alkali element resulting in equal masses $m^\sigma$ for
all the bosons. Furthermore, each species shall consist of $N_\sigma=6$ bosons
and shall experience the very same trapping potential made of a harmonic trap
superimposed with a Gaussian at the trap centre, i.e.
$U_{\sigma}(x)\equiv U(x)=x^2/2+
h/\sqrt{2\pi s^2}\;\exp(-x^2/2s^2)$ in harmonic oscillator units
$\hbar=m^\sigma=\omega=1$. We choose $h=3$ and $s=0.2$ for the
height and width of the barrier, respectively, which leads to three
bands below the barrier, each consisting of two single particle eigenstates. The
lowest band is
separated by an energy difference of $1.63$ from the first excited band and its
level spacing amounts to $\Delta E\approx 0.23$ leading to a tunneling period
of $T=2\pi/\Delta E\approx 27$ for non-interacting particles. For the contact
interaction strengths, we take $g_A (N_A-1)=0.2$, $g_B=0.75\,g_A$ and $g_{AB} =
0.05\,\sqrt{g_Ag_B}$. Furthermore, the C bosons are assumed to have no
intra-species interaction, i.e. $g_C=0$, but an attractive, vanishing or
repulsive coupling to the bosons of species A and B: $g_{XC}\equiv
g_{AC}=g_{BC}\in\{-0.5\,g_{AB},0.0,0.5\,g_{AB}\}$. Anticipating the results, we
will show that
this very weak interaction of strength $g_{XC}$ has
a significant impact on the correlation between the A and B bosons.

As the particle numbers are the same
for all species and because of the not too
different interaction strengths, we provide for each species the
same number of species states, $M_\sigma\equiv M$, and SPFs\footnote{The SPFs
are represented by
means of a harmonic discrete variable representation (DVR)
\cite{MCTDH_BJMW2000}.},
$m_\sigma\equiv m$. For preparing the initial state of the mixture, we 
block the right well by means of a high step function potential. All
bosons are then put into
the ground state of the resulting single particle Hamiltonian and, afterwards,
we let the interacting many-body system relax to its ground state by propagating
the
\mbox{ML-MCTDHB} equations of motion in imaginary time. Ramping down the
step function potential instantaneously, the resulting many-body
state
is finally propagated in real time in the original double well trap.
Afterwards, we infer the probability of a particle to be in some well and the
probability of finding two particles of the same or different species in the
same well from the corresponding reduced one-body and two-body
density matrices. 

Here we would like to point out that we do not aim at an
exhaustive study of this setup. Rather than showing a systematic parameter
scan, we would like to present one striking
example of multi-species non-equilibrium dynamics hardly being accessible in
this precision by other methods but \mbox{ML-MCTDHB}, thereby illustrating
the beneficial scaling and efficiency of the method. As we shall see, this setup
shows very interesting correlation effects. 

\subsection{Short time tunneling dynamics}
\begin{figure}
 \centering
 \includegraphics[width=0.8\textwidth]
{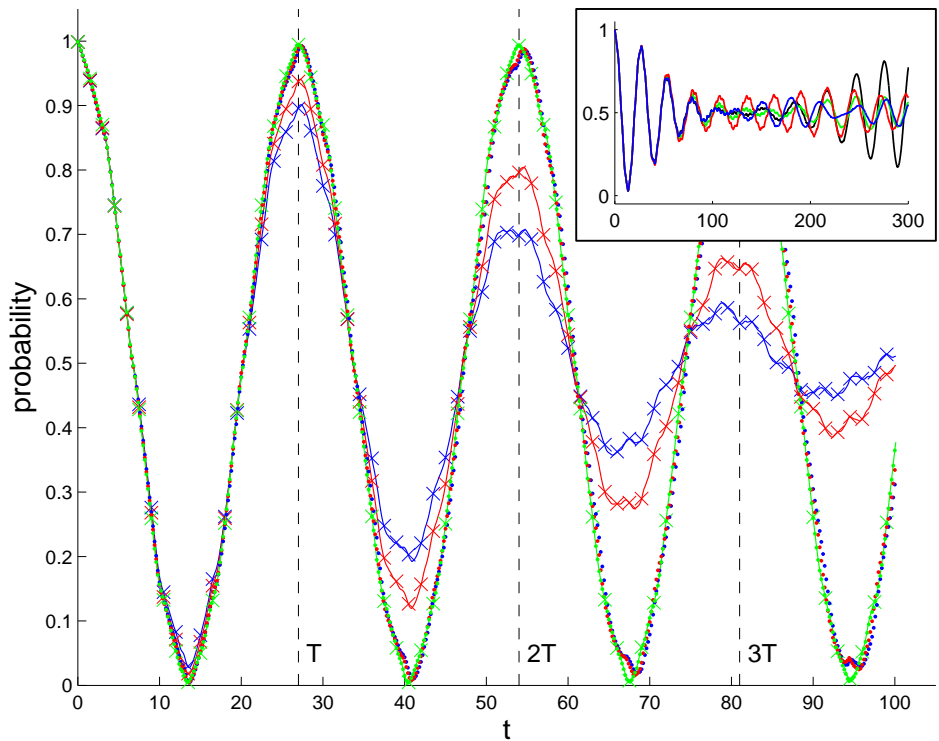}
 \caption{Tunneling dynamics of 6 A, 6 B and 6 C bosons, initially loaded into
the left well, for the interaction strengths 
$g_A (N_A-1)=0.2$, $g_B=0.75\,g_A$, $g_C=0.0$, $g_{AB} =
0.05\,\sqrt{g_Ag_B}$ and $g_{XC}\equiv g_{AC}=g_{BC}=-0.5\,g_{AB}$: The
probabilities for
finding
an A boson (blue line), a B boson (red) and a C boson (green) in the left well.
Solid lines: \mbox{ML-MCTDHB} data for $m=M=3$. Crosses: $m=M=4$. Dotted
lines: mean-field results ($m=M=1$). The first three
Rabi-tunneling periods are represented by
the dashed vertical lines. Inset: Longtime propagation of the probability to
find an A boson left ($m=3$, $M=5$) with $g_A$, $g_B$ as above. Four cases: (i)
$g_{AB}=g_{XC}=0$ (black line), $g_{AB}=0.05\,\sqrt{g_Ag_B}$ with (ii) 
$g_{XC}=0$ (green), (iii) $g_{XC}=0.5\,g_{AB}$ (red), (iv)
$g_{XC}=-0.5\,g_{AB}$ (blue).}
\label{fig_6x6x6_tunnel_prob}
\end{figure}

Let us firstly focus on the tunneling dynamics for an attractive coupling of
the C bosons to the bosons of the A and B species, i.e. $g_{XC}<0$, up to
time $t=100$. From
figure
\ref{fig_6x6x6_tunnel_prob}, we
see that the A, B and C bosons exhibit Rabi tunneling with respect to the
tunneling period on this time interval. The amplitude of the probability
oscillations, however, decreases in the course
of time for the A and the
B bosons. This decrease can be interpreted as a temporal equilibration of the
occupation probability of the left well as one can infer from the inset of
figure \ref{fig_6x6x6_tunnel_prob} showing a somewhat lower accuracy long time
propagation (see below). We also clearly see that
the decrease of the probability amplitude
is a genuine many-body property, not present in the mean-field description via
coupled Gross-Pitaevskii equations. In
contrast to this, the tunneling
amplitude of the C bosons is not damped and its dynamics in the many-body
calculation coincides with the mean-field description. This is a consequence
of the vanishing intra-species and the very weak inter-species interaction
strength. 

\begin{figure}
 \centering
 \includegraphics[width=0.8\textwidth]
{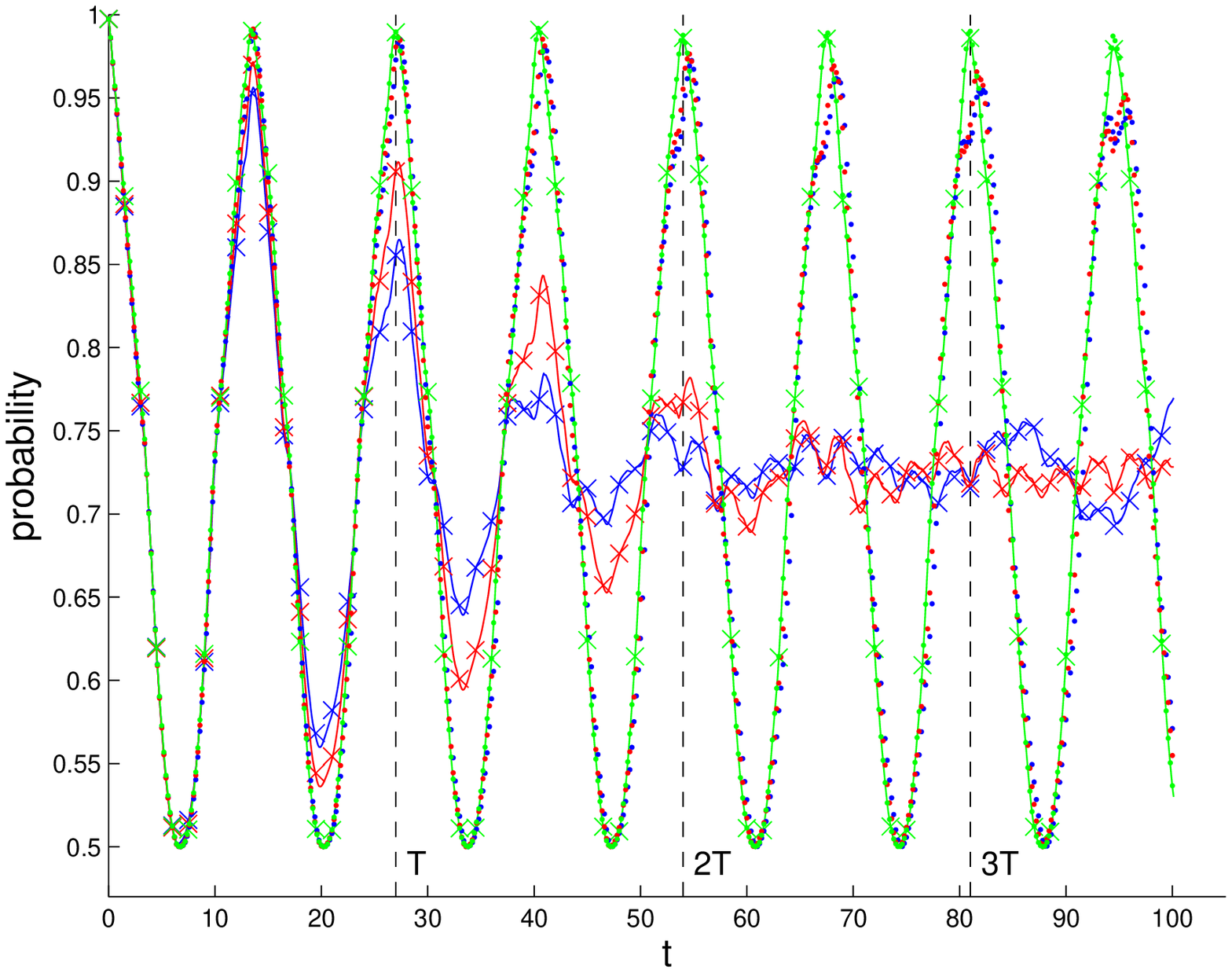}
 \caption{Time
evolution of the probabilities for finding two A bosons (blue line), two B
bosons (red), two
C bosons (green) in the same well for the setting of figure
\ref{fig_6x6x6_tunnel_prob}. Solid lines: \mbox{ML-MCTDHB} data with $m=M=3$.
Crosses: $m=M=4$. Dotted lines: mean-field results.}
\label{fig_6x6x6_tunnel_jointprob}
\end{figure}

A further phenomenon not
captured in the mean-field picture is unravelled in figure
\ref{fig_6x6x6_tunnel_jointprob}: The probabilities for finding two bosons
of the same species in the same well oscillate between 0.5 and
1.0 with the frequency $2/T$ in the mean-field calculation. In the many-body
calculation, however, the probability for finding two A or B bosons in
the same well features damped oscillations leading to a
saturation of 0.73, which indicates a bunching tendency, while the probability
of finding two C bosons stays oscillating between 0.5 and 1.0.

\begin{figure}[h]
\centering
\subfigure[]{
\centering
\includegraphics[width=0.5\textwidth]
{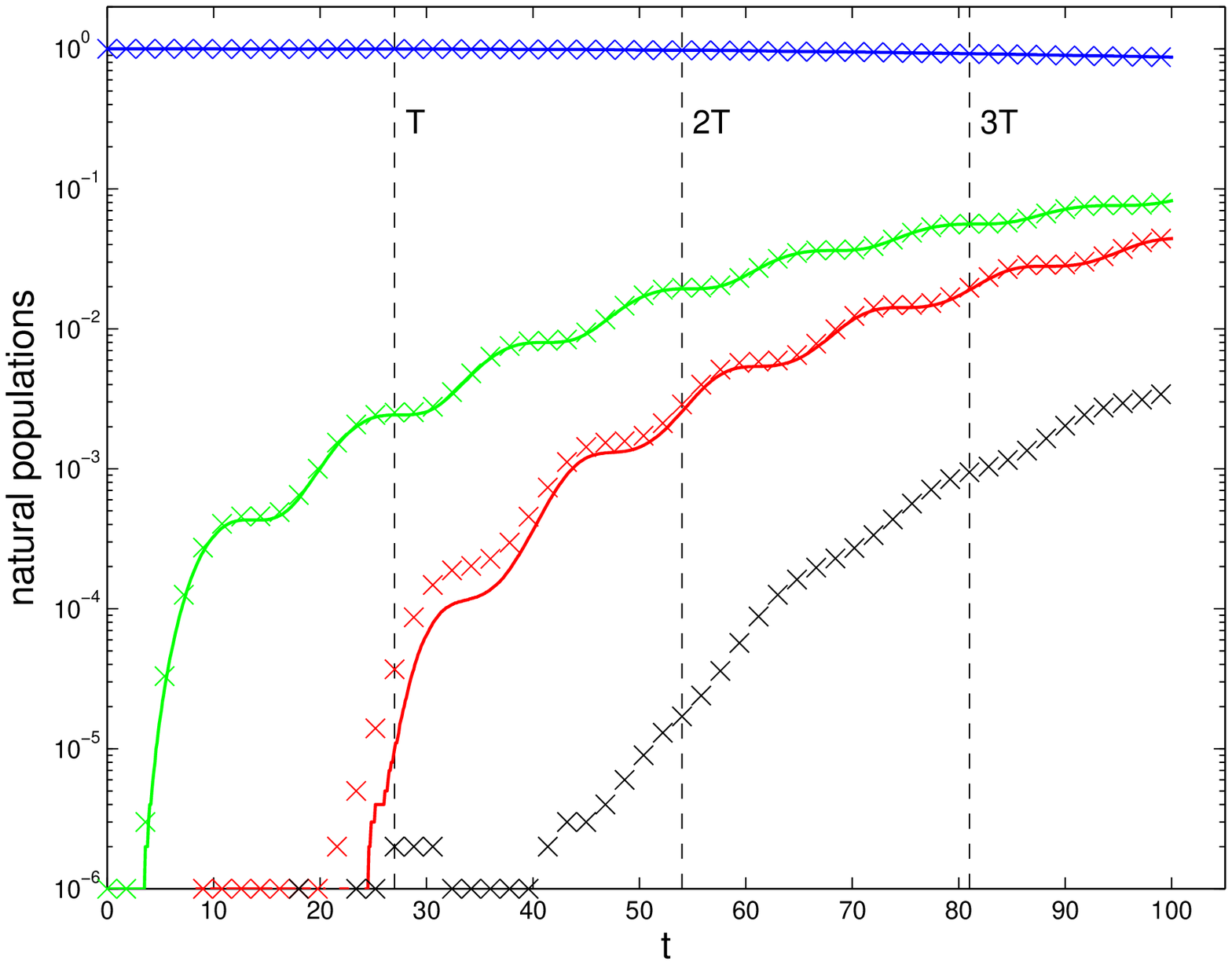}
 \label{sfig_6x6x6_natpop_spec}
}
\subfigure[]{
 \centering
 \includegraphics[width=0.5
\textwidth]{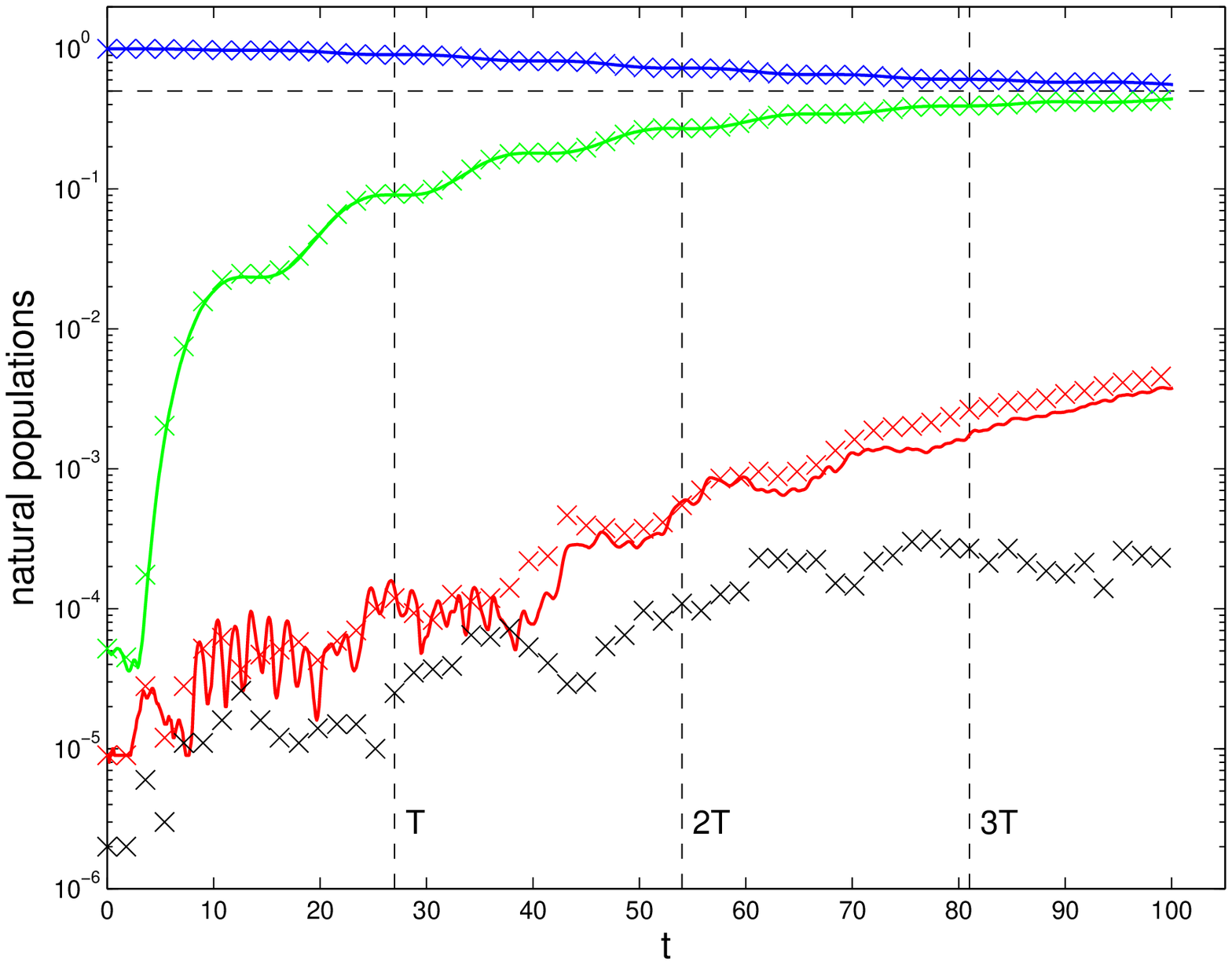}
 \label{sfig_6x6x6_natpop_partA}
}
 \caption{Time evolution of the eigenvalues of the reduced
density matrices of (a)
the species A, $[\eta_{1,A}]$, and (b) a single A boson, $[\rho_{1,A}]$,
(cf. (\ref{rho_1spec}), (\ref{rho_1part})). Solid lines: $m=M=3$
\mbox{ML-MCTDHB}
results. Crosses: $m=M=4$. The horizontal dashed line in (b)
indicates an ordinate value of 0.5, i.e. the perfect twofold fragmented state.
Parameters as in figure
\ref{fig_6x6x6_tunnel_prob}.}         
\label{fig_6x6_natpop_part}
\end{figure}

For discussing the convergence of the simulation, the \mbox{ML-MCTDHB}
calculation for $m=M=3$ is compared with the results for $m=M=4$ in
figures \ref{fig_6x6x6_tunnel_prob} and \ref{fig_6x6x6_tunnel_jointprob}.
The single particle probabilities show an excellent
agreement. Only for the joint probability of finding two particles of the same
or different (not shown)
kind in the same well, there are marginal deviations.
Hence, we can definitely regard the simulation as being converged. This
judgement is also
supported by the time evolution of the natural populations: From figure
\ref{sfig_6x6x6_natpop_spec}, we infer that most of the time only two natural
orbitals
significantly contribute to the reduced density matrix of the whole species A
and, hence, to the total wave function. For times larger
than $\sim70$, a third species state gains a weight more than 1\%. Thus,
much less species states than $M=(N+m-1)!/[N!(m - 1)!]$, i.e. the full CI
limit on the species layer, are enough for a fair representation of the total
wave function.
Figure
\ref{sfig_6x6x6_natpop_partA}
shows that the initially fully condensed state of the A bosons evolves into
a two-fold fragmented state. Increasing the
number of particle
SPFs from $m=3$ to $4$ just leads to a reshuffling of the third-highest natural
population without affecting the results. The natural populations corresponding
to a B boson show a similar behaviour due to the similar intra-species
interaction strengths (not shown). In contrast to this, the C bosons stay in a
condensed state and become depleted only by $1.2\%$ in the long time
propagation up to $t=300$ (not shown). Please note that the extra species layer
is crucial for this
convergence check: Our $m=M=4$ simulation lasted roughly a week\footnote{For
$n=250$ gridpoints on an Intel\textregistered$\,$ Xeon\textregistered$\,$  CPU
E5530  with 2.40GHz.}, while a
corresponding MCTDHB calculation would require to propagate 146 times
more coefficients.

\subsection{Long time propagation and build up of correlations}
Now let us explore the tunneling on longer time scales with a somewhat
lower
accuracy calculation choosing $m=3$ and $M=5$. A comparison with a $m=3$, $M=4$
simulation shows only very small, quantitative deviations in the observables
under consideration (plots not shown). In the inset of figure
\ref{fig_6x6x6_tunnel_prob}, the time-evolution
of the probability to find an A boson in the left well is shown comparing four
different situations, namely $g_{AB}> 0$ with $g_{XC}=0$,
$g_{XC}>0$ or $g_{XC}<0$ and
$g_{AB}=g_{XC}=0$. Although all the inter-species interaction strengths
are much smaller than $g_A$ and $g_B$, their concrete values have a
strong influence on the tunneling dynamics: For no inter-species interactions,
there happens to be a partial revival of the tunneling oscillation after a
temporal equilibration. In the case of $g_{AB}>0$, only for a
vanishing or attractive coupling of the C species to the other species one can
observe such a temporal equilibration state with subsequent partial
tunneling revival of, however, smaller amplitude in comparison to the
former case. A repulsive coupling between the C and the other species does
not
result in a complete temporal equilibration to a probability of $0.5$ but rather
leads to a reduction of the amplitude of the oscillations around the
equilibration value to $0.12$. While the B bosons show a
dynamics similar to the A bosons, the C bosons tunnel almost unaffected by the
inter-species interactions and exhibit mean-field tunneling
oscillations (plots not shown).

In order to measure the correlations between different
species, we compare the conditional probability of finding a $\sigma$ boson in
e.g. the left well given that a $\sigma'$ boson has already been found there
with the marginal probability of finding a $\sigma$ boson in the left well:
Let $P(\sigma, L;\sigma', L)$ denote the probability for finding a $\sigma$ and
a $\sigma'$ boson in
the left well and let $P(\sigma, L)$ ($P(\sigma', L)$) be the probability for
finding a $\sigma$ ($\sigma'$) boson in the left well. Then the above mentioned
correlation measure reads $f_{LL}(\sigma,\sigma'):=P(\sigma, L;\sigma', L)
/[P(\sigma, L)P(\sigma', L)]$ and we similarly define $f_{RR}(\sigma,\sigma')$
for the right well. Please note that $(f_{LL},f_{RR})$ is a
straightforward extension of the
diagonal elements of the $g_2$ coherence / correlation measure
\cite{quantum_theory_of_opt_coherence_Glauber1963} to spatially discrete systems
with
distinguishable components. The dynamics of the centre of mass positions of the
$\sigma$ and the $\sigma'$
species has an impact on $f_{LL}$ and $f_{RR}$, of course.
In order to diminish this impact, we
finally construct our correlation measure for finding a $\sigma$ and a $\sigma'$
boson in the same well as
$f(\sigma,\sigma'):=([f_{LL}(\sigma,\sigma')^2+f_{RR}(\sigma,\sigma')^2]
/2)^{1/2}$. If the $\sigma$ and the $\sigma'$ bosons tunnel independently,
$f(\sigma,\sigma')$ will be unity. A value of $f(\sigma,\sigma')$ greater
(smaller) than one indicates an overall bunching (anti-bunching) tendency.

\begin{figure}
 \centering
 \includegraphics[width=0.8\textwidth]
{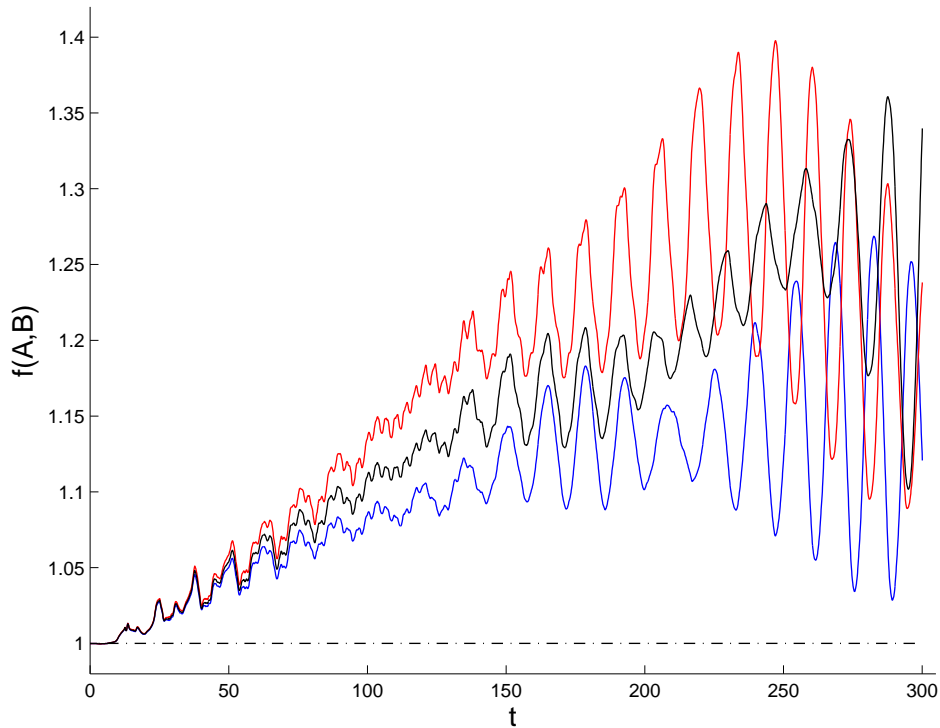}
 \caption{Time evolution of the inter-species correlation measure $f(A,B)$
for $g_{AB}=0.05\,\sqrt{g_Ag_B}$ and  $g_{XC}=0$ (black line),
$g_{XC}=0.5\,g_{AB}$ (red) and $g_{XC}=-0.5\,g_{AB}$ (blue). $m=3$, $M=5$.
Corresponding to
absent inter-species interactions, the dash-dotted
serves as a reference. Other parameters as in figure
\ref{fig_6x6x6_tunnel_prob}.}
\label{fig_corr}
\end{figure}

In figure \ref{fig_corr}, we find that a bunching tendency between an A and a B
boson clearly builds up with a maximal correlation measure $f(A,B)$ of up to
$40\%$ above unity. For  $75\lesssim t \lesssim
225$, this bunching tendency turns out to be most intense for
the repulsive coupling of the C bosons to the other species and becomes least
intense for an attractive coupling. In the absence of inter-species interactions
with the C species, i.e. if $g_{AB}>0$ is the only non-vanishing inter-species
interaction strength, the correlation measure $f(A,B)$ lies mostly inbetween
these two curves. For a large
fraction of the propagation time, the coupling of the C bosons to the other
two species can thus control the inter-species correlations between species A
and B up to a certain degree. Due to the fact that the C bosons
approximately
perform Rabi-oscillations with respect to the occupation probability of the
left well, one might come to the conclusion that the C bosons provide a
time-dependent potential for the other two species hardly experiencing
a backaction on the considered time-scale. That this descriptive
picture can only be approximately valid up to a certain time, can be inferred
from the second largest natural population of $[\eta_{1,C}]$, which monotonously
increases up to $6.2\%$ ($g_{XC}<0$) and $14\%$ ($g_{XC}>0$), respectively.
%%%%%%%%%%%%%%%%%%%%%%%%%%%%%%%%%%%%%%%%%%%%%%%%%%%%%%%%%%%%%%%%%%%%%%%%%%%%%%%%
%%%%%%%%%%%%%%%%%%%%%%%%%%%%%%%%%%%%%%%%%%%%%%%%%%%%%%%%%%%%%%%%%%%%%%%%%%%%%%%%
\section{Conclusion and outlook}\label{sec_concl}
We have presented a novel {\it ab initio} method for
simulating the non-equilibrium dynamics of mixtures of ultra-cold bosons. In
particular, ML-MCTDHB is suitable for dealing with explicitly time-dependent
systems, which will be explored in future works. Being based on an expansion in
terms of permanents and on a multi-layer
ansatz, our \mbox{ML-MCTDHB} method takes optimally and efficiently the bosonic
exchange symmetry within each species into account and allows for adapting the
ansatz to
system specific intra- and inter-species correlations. Hereby, the numbers of
provided single particle functions and species states serve as control
parameters for ensuring convergence. For any choice of these numbers of basis
functions, \mbox{ML-MCTDHB} rotates the species states and single particle
functions such that one obtains a variationally optimal representation of the
many-body wave function at any instant in time. This allows to achieve
convergence with a much smaller basis than methods being based on a
time-independent basis. Moreover, if the inter-species interactions are not too
strong, i.e. do not require to consider as many species states as there are
number state configurations for a given number of single particle functions,
ML-MCTDHB proves to have a much better scaling than the best state-of-the-art
method MCTDHB
\cite{
role_of_excited_states_in_splitting_of_BEC_by_time_dep_barrier_Steltsov_PRL2007,
MCTDHB_many_body_dynacmics_of_bosonic_systems_phys_rev_a_Alon_Streltsov_Cederbau
m, MCTDHX_recursively_Cederbaum2012}.
In the case of only a single species state and one
single particle function, \mbox{ML-MCTDHB}
reduces to coupled Gross-Pitaevskii equations.

Employing ML-MCTDHB for a tunneling scenario of three species, we have entered
a parameter regime which is hardly accessible by other methods in such a
controlled precision. Our simulations show that the imbalances of the
populations can feature a temporal equilibration with subsequent revival of the
population oscillations, where the duration of and the fluctuations around the
equilibration state as well as the degree of completeness of the revival
crucially depend on the inter-species interaction strengths. In our setup, we
have furthermore found two-body bunching correlations between
the first two species. The strength of this correlation can be tuned by a weak
attractive or repulsive coupling of the third species - with no intra-species
interaction - to the first two species without significantly altering the
tunneling dynamics of that third species.

In this paper, \mbox{ML-MCTDHB} has been formulated for systems confined by
quasi-one-dimensional traps and interacting via contact interactions. A
direct generalization to arbitrary dimensions and interaction potentials is
possible, of course. Moreover, it is also feasible to generalize
\mbox{ML-MCTDHB} further by applying the multi-layering concept on the level of
the single particle functions, which allows for optimally describing bosons in
quasi-one- or -two-dimensional traps embedded in three-dimensional space with or
without internal degrees of freedom \cite{ml-mctdhb_unpub}.
Incorporating internal degrees of freedom on the level of SPFs then allows
for taking particle converting interactions into account.
%%%%%%%%%%%%%%%%%%%%%%%%%%%%%%%%%%%%%%%%%%%%%%%%%%%%%%%%%%%%%%%%%%%%%%%%%%%%%%%%
%%%%%%%%%%%%%%%%%%%%%%%%%%%%%%%%%%%%%%%%%%%%%%%%%%%%%%%%%%%%%%%%%%%%%%%%%%%%%%%%
\ack
The authors would like to thank Hans-Dieter Meyer and Jan Stockhofe for
fruitful discussions on MCTDH methods and symmetry conservation.
Particularly, the authors would like to thank Jan Stockhofe for the DVR
implementation of the \mbox{ML-MCTDHB}
code. S.K. gratefully acknowledges financial support by the Studienstiftung des
deutschen Volkes. L.C. and P.S. gratefully acknowledge funding by the
Deutsche Forschungsgemeinschaft in the framework of the SFB 925 ``Light induced
dynamics and control of correlated quantum systems''.
%%%%%%%%%%%%%%%%%%%%%%%%%%%%%%%%%%%%%%%%%%%%%%%%%%%%%%%%%%%%%%%%%%%%%%%%%%%%%%%%
%%%%%%%%%%%%%%%%%%%%%%%%%%%%%%%%%%%%%%%%%%%%%%%%%%%%%%%%%%%%%%%%%%%%%%%%%%%%%%%%
\appendix
\section{Ingredients for the evolution equations of
the species states}\label{app_spec_layer}
\setcounter{section}{1}
The matrix elements entering (\ref{eqm_spec}) read:
\begin{eqnarray}
 {[h_\sigma]}_{jk} &=& \langle\phi^{(\sigma)}_j| 
  [\frac{\hat p_\sigma^2}{2m^\sigma}+
	  U_\sigma(\hat x_\sigma)] |\phi^{(\sigma)}_k\rangle,\\
 {[v_{\sigma}]}_{jkqp} &=& g_\sigma
\langle\phi^{(\sigma)}_j\phi^{(\sigma)}_k|\delta(\hat x_1^\sigma-
		\hat x_2^\sigma)|\phi^{(\sigma)}_q\phi^{
(\sigma)}_p\rangle,\\
{[w_{\sigma\sigma'}]}^{jk}_{uv} &=&g_{\sigma\sigma'}\sum_{\vec l|N_{\sigma'}-1}
\sum_{q,p=1}^{m_{\sigma'}}
\langle\phi^{(\sigma)}_j\phi^{(\sigma')}_q|\delta(\hat x_1^\sigma-
		\hat x_2^{\sigma'})|\phi^{(\sigma)}_k\phi^{
(\sigma')}_p \rangle \times\\\nonumber
&\phantom{=}&\qquad\qquad\qquad\qquad \times Q_{\vec
l\,}(q,p)\;\left(C^{\sigma'}_{u;\vec l + \hat q}\right)^*
  C^{\sigma'}_{v;\vec l + \hat p},
\end{eqnarray}
where $Q_{\vec l\,}(q,p) =\sqrt{(l_q+1)(l_p+1)}$. ``$\vec l|N_{\sigma'}-1$''
refers to summation over all occupation numbers summing up to $N_{\sigma'}-1$,
and $\hat q$ represents an occupation number vector with vanishing entries
except for the $q$-component being set to one. The reduced density matrix
corresponding to the species $\sigma$ can be calculated as:
\begin{equation}\label{rho_1spec}
 {[\eta_{1,\sigma}]}_{is} = \sum_{J^\sigma}
\left(A_{J^\sigma_i}\right)^*A_{J^\sigma_s},
\end{equation}
where the summation runs over all indices except for the $\sigma$ index, which
is fixed to be $r$ in the multi-index $J^\sigma_r$. For inversion,
$[\eta_{1,\sigma}]$ has to be regularized \cite{MCTDH_BJMW2000}. In analogy, the
reduced density matrix of the subsystem constituted by the $\sigma$ and
$\sigma'$ species is given as:
\begin{equation}\label{rho_2spec}
 {[ \eta_{2, \sigma\sigma'}]}_{sutv}= \sum_{J^{\sigma\sigma'}}
\left(A_{J^{\sigma\sigma'}_{su}}\right)^*A_{J^{\sigma\sigma'}_{tv}},
\end{equation}
where $\sigma\neq\sigma'$.
Here, the summation runs over all indices except for the $\sigma$ and $\sigma'$
index, which are fixed to be $r$ and $q$ in $J^{\sigma\sigma'}_{rq}$.

\section{Ingredients for the evolution equations of
the SPFs}\label{app_part_layer}
In the particle layer, the mean-field operator matrices for the intra- and
inter-species interaction are given as: 
\begin{eqnarray}
 {[ \hat v_{\sigma}]}_{kp} &=&  g_\sigma\int \textup{d}x\,
\left(\phi^{(\sigma)}_k(x)
			    \right)^*
			    \phi^{(\sigma)}_p(x)\,\delta(x-\hat x_\sigma),\\
 {[ \hat w_{\sigma\sigma'}]}_{kp} &=&  g_{\sigma\sigma'}\int \textup{d}x\,
			    \left(\phi^{(\sigma')}_k(x)\right)^*
			    \phi^{(\sigma')}_p(x)\,\delta(x-\hat x_\sigma).
\end{eqnarray}
The one-body reduced density matrix of a $\sigma$ boson, which also has to
be regularized \cite{MCTDH_BJMW2000}, can be calculated as:
\begin{equation}\label{rho_1part}
 {[\rho_{1,\sigma}]}_{ij} = \frac{1}{N_\sigma} \sum_{u,v=1}^{M_\sigma}
    {[\eta_{1,\sigma}]}_{uv} \sum_{\vec l|N_\sigma-1} Q_{\vec l\,}(i,j)\,
    \left(C^{\sigma}_{u;\vec l+\hat i}\right)^*C^{\sigma}_{v;\vec l+\hat j}.
\end{equation}
For the reduced density matrices of two $\sigma$ bosons and of a $\sigma$ and a
$\sigma'$ boson ($\sigma\neq\sigma'$), one has the following expressions:
\begin{eqnarray}\fl\label{rho_2Apart}
 {[ \rho_{ 2, \sigma\sigma}]}_{jkqp} = \frac{1}{N_\sigma}
\sum_{u,v=1}^{M_\sigma}
    {[\eta_{1,\sigma}]}_{uv} \sum_{\vec l|N_\sigma-2}
    P_{\vec l\,}(j,k)P_{\vec l\,}(q,p)\,
  \left(C^{\sigma}_{u;\vec l+\hat j+\hat k}\right)^*
  C^{\sigma}_{v;\vec l+\hat q +\hat p},\\\fl\label{rho_1A1Bpart}
 {[ \rho_{ 2, \sigma\sigma'}]}_{jkqp} =  \frac{1}{N_\sigma}
  \sum_{s,t=1}^{M_\sigma} \sum_{u,v=1}^{M_{\sigma'}} {[ \eta_{ 2,
\sigma\sigma'}]}_{sutv} \sum_{\vec l|N_\sigma-1} 
\sum_{\vec m|N_{\sigma'}-1}Q_{\vec l\,}(j,q)Q_{\vec m\,}(k,p) \times\\\nonumber
  \qquad\qquad\qquad\qquad\qquad\qquad\qquad\times \left(C^{\sigma}_{s;\vec
l+\hat j}\right)^*
  C^{\sigma}_{t;\vec l+\hat q} \left(C^{\sigma'}_{u;\vec m+\hat k}\right)^*
  C^{\sigma'}_{v;\vec m+\hat p},
\end{eqnarray}
with $P_{\vec l\,}(j,p)=\sqrt{(l_j+\delta_{jp}+1)(l_p+1)}$ and $\delta_{jp}$
denoting the Kronecker delta function. We remark that the ML-MCTDHB
code employs a different strategy than MCTDHB
\cite{general_mapping} for
evaluating the various density matrices and the action of annihilation and
creation operators on number states (cf. (\ref{eqm_spec})) and refer to
\cite{ml-mctdhb_unpub} for the details.

%%%%%%%%%%%%%%%%%%%%%%%%%%%%%%%%%%%%%%%%%%%%%%%%%%%%%%%%%%%%%%%%%%%%%%%%%%%%%%%%
%%%%%%%%%%%%%%%%%%%%%%%%%%%%%%%%%%%%%%%%%%%%%%%%%%%%%%%%%%%%%%%%%%%%%%%%%%%%%%%%
\section*{References}
\bibliography{ml-b_refs_iop_style}
\bibliographystyle{unsrt}

\end{document}